\def\year{2022}\relax
\documentclass[letterpaper]{article} 
\usepackage{aaai22}  
\usepackage{times}  
\usepackage{helvet}  
\usepackage{courier}  
\usepackage[hyphens]{url}  
\usepackage{graphicx} 
\usepackage{natbib}  
\usepackage{caption} 
\DeclareCaptionStyle{ruled}{labelfont=normalfont,labelsep=colon,strut=off} 
\usepackage{algorithm}
\usepackage[table,xcdraw]{xcolor}
\usepackage[hang,flushmargin]{footmisc} 
\usepackage{booktabs}
\usepackage{amsfonts}
\usepackage{amssymb} 
\usepackage{bm,bbm}
\usepackage{hyperref}
\usepackage[noend]{algpseudocode}
\newcommand\Algphase[1]{%
\vspace*{-0.5\baselineskip}\Statex\hspace*{\dimexpr-\algorithmicindent-2pt\relax}\rule{8.56cm}{0.4pt}%
\Statex\hspace*{-\algorithmicindent}\textbf{#1}%
\vspace*{-0.5\baselineskip}\Statex\hspace*{\dimexpr-\algorithmicindent-2pt\relax}\rule{8.56cm}{0.4pt}%
}
\definecolor{dblue}{RGB}{0,0,255}

\usepackage{newfloat}
\usepackage{listings}
\floatstyle{ruled}
\newfloat{listing}{tb}{lst}{}
\floatname{listing}{Listing}

\title{SSAST: Self-Supervised Audio Spectrogram Transformer}
\author{
   Yuan Gong, Cheng-I Jeff Lai, Yu-An Chung, James Glass
}
\affiliations{
    MIT Computer Science and Artificial Intelligence Laboratory, Cambridge, MA 02139 \\
    \texttt{\{yuangong, clai24, andyyuan, glass$\}$@mit.edu}
}

\begin{document}

\maketitle

\begin{abstract}

Recently, neural networks based purely on self-attention, such as the Vision Transformer~(ViT), have been shown to outperform deep learning models constructed with convolutional neural networks (CNNs) on various vision tasks, thus extending the success of Transformers, which were originally developed for language processing, to the vision domain.
A recent study~\cite{gong21b_interspeech} showed that a similar methodology can also be applied to the audio domain. Specifically, the Audio Spectrogram Transformer~(AST) achieves state-of-the-art results on various audio classification benchmarks. However, pure Transformer models tend to require more training data compared to CNNs, and the success of the AST relies on supervised pretraining that requires a large amount of labeled data and a complex training pipeline, thus limiting the practical usage of AST.

This paper focuses on audio and speech classification, and aims to reduce the need for large amounts of labeled data for the AST by leveraging self-supervised learning using unlabeled data. Specifically, we propose to pretrain the AST model with joint discriminative and generative masked spectrogram patch modeling (MSPM) using unlabeled audio from AudioSet and Librispeech. We evaluate our pretrained models on both audio and speech classification tasks including audio event classification, keyword spotting, emotion recognition, and speaker identification. The proposed self-supervised framework significantly boosts AST performance on all tasks, with an average improvement of 60.9\%, leading to similar or even better results than a supervised pretrained AST. To the best of our knowledge, it is the first \emph{patch-based} self-supervised learning framework in the audio and speech domain, and also the first self-supervised learning framework for AST. Code at \href{https://github.com/YuanGongND/ssast}{\color{blue}{https://github.com/YuanGongND/ssast}}.



\end{abstract}

\section{Introduction}
Pure self-attention based deep learning architectures, such as the Vision Transformer~\cite{dosovitskiy2021image} and its variants (e.g., DeiT~\cite{touvron2020deit}, T2T-ViT~\cite{yuan2021tokens}) have been shown to outperform CNN models~\cite{lecun1995convolutional} of similar size on various vision tasks. Such models differ from CNN models or CNN-attention hybrid models in that they do not contain non-degenerated convolutions~\cite{chen2021empirical} and thus have less inductive bias such as spatial locality or translation equivariance, and are more data-driven. In the audio and speech domain, the recently proposed Audio Spectrogram Transformer (AST)~\cite{gong21b_interspeech} and the Keyword Transformer~\cite{berg21_interspeech} also achieve new state-of-the-art performance on audio scene classification and keyword spotting. Despite the strong performance, a critical issue of such pure self-attention based models is they tend to require more training data than CNNs~\cite{dosovitskiy2021image}.  For example, the ViT outperforms CNNs only when the training data volume is larger than about 100 million samples. AST also does not perform well when it is trained from scratch, and the success of AST strongly relies on supervised pretraining. Since labeled speech and audio data is limited, AST uses cross-modal pretraining with ImageNet data~\cite{deng2009imagenet}. 
However, in practice, supervised pretraining on ImageNet data is complex~\cite{he2019bag} and expensive, and also constrains the vision and audio models to have a similar architecture and use the same patch size and shape. Further, the validity and transferability of such cross-modal pretraining for a specific audio or speech task are unclear.

While annotating audio and speech data is expensive, we can easily get web-scale unlabeled audio and speech data from radio or YouTube. This motivates us to explore Self-Supervised AST (SSAST) that leverages \emph{unlabeled} data to alleviate the data requirement problem. In this paper, we present a novel joint discriminative and generative Masked Spectrogram Patch Modeling (MSPM) based self-supervised learning (SSL) framework that can significantly improve AST performance with limited labeled data.  Previous self-supervised learning methods such as wav2vec~\cite{schneider2019wav2vec} or autoregressive predictive coding (APC)~\cite{chung2019unsupervised} use an objective that predicts future or masked temporal spectrogram frames, thus potentially learning only the temporal structure of the spectrogram. In contrast, the objective of MSPM is to predict a specific frequency band in a specific time range (i.e., a ``spectrogram patch'') given the neighboring band and time information, which allows the model to learn both the temporal and frequency structure. The spectrogram patch can be an arbitrary shape and size, e.g., it can be a conventional time frame or a square patch.

In addition, most previous SSL research considers either only speech or only audio events, but in this work, we show that the SSL model can be generalized to both speech and audio tasks. Specifically, we pretrain our model using both Librispeech and AudioSet, and evaluate the model on a variety of speech and audio tasks including audio event classification, keyword spotting, speaker identification, and speech emotion recognition. Our experiments demonstrate the effectiveness of the proposed MSPM framework: a model pretrained with MSPM can significantly outperform from-scratch models for all 6 benchmarks we evaluated with an average improvement of $60.9\%$, and the performance can even match or outperform supervised pretrained models. The contributions of this work are two-fold:

\begin{enumerate}
    \item We propose MSPM, a novel patch-based joint discriminative and generative self-supervised learning framework. With MSPM pretraining, our SSAST model matches or outperforms previous supervised pretrained AST. To the best of our knowledge, MSPM is the first \emph{patch-based} self-supervised learning framework in the audio and speech domains, and SSAST is the first self-supervised pure self-attention based audio classification model. Further, we conduct extensive experiments to thoroughly investigate the design choices and quantify the performance impact of each factor.
    \item We show that pretraining with both speech and audio datasets noticeably improves the models' generalization ability, and leads to better performance than pretraining with dataset from a single domain. As a consequence, our SSAST model performs well on both speech and audio downstream tasks. Previous work typically only uses datasets in a single domain for pretraining.
\end{enumerate}

\section{Self-Supervised Audio Spectrogram Transformer}
\label{sec:method}

In this section, we first review the AST architecture and then discuss the proposed joint discriminative and generative masked spectrogram patch prediction (MSPM) self-supervised learning framework, and the design details.

\begin{figure}[!t]
    \centering
    \includegraphics[width=6.75cm]{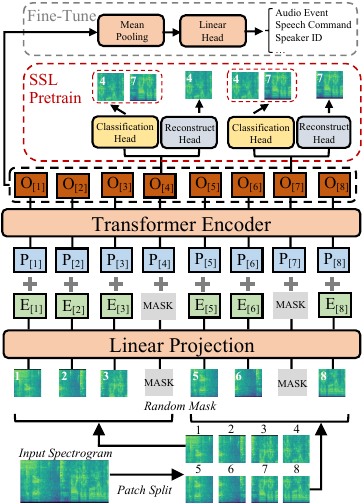}
    \caption{The proposed self-supervised AST. The 2D audio spectrogram is split into a sequence of 16$\times$16 patches without overlap, and then linearly projected to a sequence of 1-D patch embeddings $E$. Each patch embedding is added with a learnable positional embedding $P$ and then input to the Transformer encoder. The output of the Transformer $O$ is used as the spectrogram patch representation. During self-supervised pretraining, we randomly mask a portion of spectrogram patches and ask the model to 1) find the correct patch at each masked position from all masked patches; and 2) reconstruct the masked patch. The two pretext tasks aim to force the AST model to learn both the temporal and frequency structure of the audio data. During fine-tuning, we apply a mean pooling over all patch representation $\{O\}$ and use a linear head for classification.}
    \label{fig:ilus}
\end{figure}

\subsection{AST Model Architecture}
\label{sec:arc}

As shown in Figure~\ref{fig:ilus}, we intentionally follow as close as possible to the original AST architecture to make a fair performance comparison. First, the input audio waveform of~$t$ seconds is converted into a sequence of~128-dimensional log Mel filterbank~(fbank) features computed with a~25ms Hanning window every~10ms. This results in a~$128\times100t$ spectrogram as input to the AST. We then split the spectrogram into a sequence of $16\times16$ patches. We flatten each~$16\times16$ patch to a 1D 768-dimensional patch embedding with a linear projection layer. We refer to this linear projection layer as the patch embedding layer and the output as patch embedding $E$.
Since the Transformer architecture does not capture the input order information and the patch sequence is also not in temporal order, we add a trainable positional embedding~(also of size~768) $P$ to each patch embedding to allow the model to capture the spatial structure of the 2D audio spectrogram. The resulting sequence is then input to the Transformer. A Transformer consists of several encoder and decoder layers.
Since the AST is designed for classification tasks, we only use the encoder of the Transformer that has an embedding dimension of~768, 12 layers, and 12 heads, which are the same as those in original AST~\cite{gong21b_interspeech}. We refer to the output of the Transformer encoder as patch representation $O$. During fine-tuning and inference, we apply a mean pooling over the sequence of patch representation $\{O\}$ to get the audio clip level representation, and then use a linear head for classification. 

While we aim to follow the architecture of the original AST, we made two modifications for self-supervised learning. First, in the original AST, a \texttt{[CLS]} token is appended to the beginning of the input sequence of the Transformer encoder, and the output representation of the \texttt{[CLS]} token is used as the audio clip level representation.  In this work, we apply mean pooling over all patch representation $\{O\}$ as the audio clip level representation. This is because the original AST uses supervised pretraining and the supervision is applied on the \texttt{[CLS]} token, thus the output representation of the \texttt{[CLS]} learns to summarize the entire sequence during pretraining and can be used as audio clip level representation. In contrast, for our self-supervised pretraining framework, supervision is applied to each individual patch representation, and the mean of all patch representations is a better summary of the audio clip. Second, in the original AST, spectrogram patches are split with overlap, and the overlap was shown to improve model performance. In this work, we split the patch without overlap during pretraining to not allow the model to use overlapped edges as a shortcut for the task prediction instead of learning a meaningful representation. In the fine-tuning and inference steps, we split the patch with an overlap of 6 in the same fashion as the original AST. 

While we pretrain the model using fixed-length audio data (10 seconds), AST supports variable length input by simply interpolating or truncating the positional embedding to the downstream task audio length. 

\begin{figure}[!t]
    \centering
    \includegraphics[width=8.2cm]{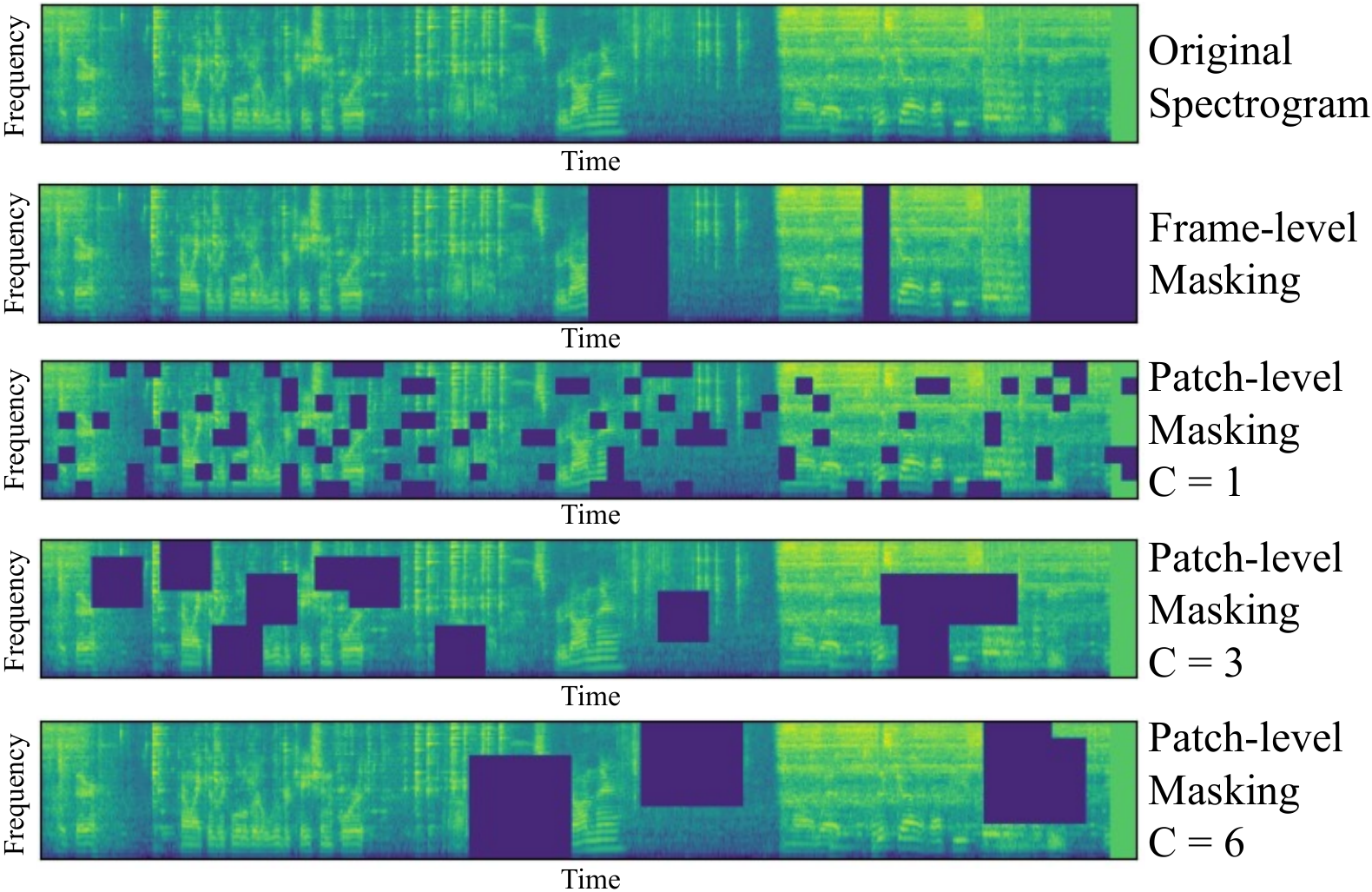}
    \caption{Illustration of the proposed patch-level masking with different cluster factor $C$ but same total masking area. The model is forced to learn more global spectrogram structure with a larger $C$, and more local structure with a smaller $C$. To make the model learn both local and global structure, we use random $C$ during pretraining. Compared with frame-level masking SSL methods that potentially only learn temporal frame structure, patch-based masking allows the model to learn both temporal and frequency spectrogram structure.}
    \label{fig:masking}
\end{figure}

\subsection{Joint Discriminative and Generative Masked Spectrogram Patch Modeling}

In this section, we introduce the proposed self-supervised pretraining framework. We first show our masking strategy and then discuss the pretext task (i.e., the self-supervised learning task in the pretraining stage) in detail.

\begin{algorithm}[!t]
\caption{Joint Discriminative and Generative Masked Spectrogram Patch Modeling}
\label{alg:pretrain}
\begin{algorithmic}[1]
\Require{
\quad \newline
Unlabeled Audio Dataset $\mathcal{D}$, AST Model $\mathcal{M}$}
\Algphase{SampleMaskIndex ($N,C$)}
$\triangleright$ Randomly sample patches to mask \newline
\textbf{Input:} $\#$Masked Patches $N$; Cluster Factor $C$ \newline
\textbf{Output:}  Masked Patch Position Index Set $I$
\While {$|I|<N$}
\State draw index $i\sim unif \{1,512\}$ 
\State get set $I_c$ = $[C^2$-1 indexes neighboring $i, i]$
\State $I=I\cup I_c$
\EndWhile
\State $I=I[1:N]$ $\triangleright$ Guarantee to mask exactly $N$ patches \newline
\Return  $I$

\Algphase{MSPM ($\mathcal{D},\mathcal{M}$)}
\textbf{Input:} $\mathcal{D},\mathcal{M}$, Number of Masked Patches $N$
\For{every epoch}
\For{$X \in \mathcal{D}$}
\State split $X$ into 512 patches $x=\{x_1, x_2, ..., x_{512}\}$
\State $E=\mathcal{M}_{patch embedding}(x)$
\State draw $C\sim unif \{3,6\}$ 
\State $I=$ SampleMaskIndex($C,N$)
\State $E_I$ = $E_{mask}$  \quad $\triangleright$ Mask the Patch Embeddings
\State $O=\mathcal{M}_{transformer}(E+P)$
\State $\mathcal{L}_d=0,\mathcal{L}_g=0$
\For{$i\in I$}
\State $r_i=\mathcal{M}_{reconstruction\_head}(O_i)$
\State $c_i=\mathcal{M}_{classification\_head}(O_i)$
\State $\mathcal{L} \mathrel{+}= \mathcal{L}_d(x_i,c_i,x_I) + \lambda\mathcal{L}_g(x_i,r_i)$
\EndFor
\State$\mathcal{L} = \mathcal{L}\mathbin{/}N$
\State update $\mathcal{M}$ to minimize $\mathcal{L}$ 
\EndFor
\EndFor 
\Return $\mathcal{M}$
\end{algorithmic}
\end{algorithm}

\subsubsection{Masked Patch Sampling}

As mentioned above, during pretraining, we use a fixed-length audio of 10s and convert it to spectrogram of size $1024\times128$. AST splits the spectrogram into 512 $16\times16$ patches (8 in the frequency dimension and 64 in the time dimension). Thanks to this special design of AST, we are able to mask spectrogram patches rather than the entire time frames during pretraining, which allows the model to learn both the temporal and frequency structure of the data. In addition, as shown in Figure~\ref{fig:masking}, we use a \emph{cluster factor} $C$ to control how masked patches cluster. Specifically, we first randomly select a patch, and then mask the square centered at the patch with a side length of $C$, e.g., if $C=3$, we mask a cluster of 9 patches that has a total size of $48\times48$. The model is forced to learn more global spectrogram structure with a larger $C$, and more local structure with a smaller $C$. To make the model learn both local and global structure, we use random $C\sim[3,5]$ during pretraining. We show the details in Algorithm~\ref{alg:pretrain} line 1-5. Note that while we mainly focus on using $16\times16$ patches in this paper, MSPM actually supports patches of arbitrary size and shape.

\subsubsection{Joint Discriminative and Generative Masked Spectrogram Patch Modeling}

As opposed to prior work that either used discriminative (e.g., wav2vec) or generative training objectives (e.g., APC), in this work, we propose to use a joint discriminative and generative objective for pretraining.

As shown in Algorithm~\ref{alg:pretrain}, each input spectrogram $X$ is split into 512 patches $x$ converted to corresponding patch embeddings $E$ (line 8-9). We then randomly generate a set $I$ of $N$ masked patch position indexes as previously described (line 10-11). For each patch that needs to be masked, we replace its patch embedding with a learnable mask embedding $E_{mask}$ (line 12). We add positional embeddings to the patch embeddings and input them to the Transformer encoder (line 13). For each masked patch $x_i$, we get the corresponding Transformer encoder output $O_i$. We then input $O_i$ to a classification head and a reconstruction head and get output $c_i$ and $r_i$, respectively (line 16-17). Both the classification and reconstruction heads are two-layer MLPs that map $O_i$ (768) to the same dimension as $x_i$ (256). We expect $r_i$ to be close to $x_i$, and the model can match correct ($x_i, c_i$) pairs. Therefore, we use the InfoNCE loss~\cite{oord2018representation} $\mathcal{L}_d$ for the discriminative objective and mean square error (MSE) loss $\mathcal{L}_g$ for the generative objective:

\begin{equation}
\mathcal{L}_d=-\frac{1}{N}\sum_{i=1}^Nlog(\frac{exp(c_i^Tx_i)}{\sum_{j=1}^Nexp(c_i^Tx_j)})
\end{equation}

\begin{equation}
\mathcal{L}_g=\frac{1}{N}\sum_{i=1}^N(r_i-x_i)^2
\end{equation}

Where $N$ is the number of masked patches. We then sum up $\mathcal{L}_d$ and $\mathcal{L}_g$ with a weight $\lambda$. In this work, we set $\lambda=10$.

\begin{equation}
\mathcal{L}=\mathcal{L}_d + \lambda\mathcal{L}_g
\end{equation}

Finally, we update the weights of the AST model $\mathcal{M}$ to minimize $\mathcal{L}$ with the optimizer (line 19-20). Note that for the discriminative task, the negative samples are sampled from the \emph{same} spectrogram, i.e., the model aims to pick the correct patch for each masked position from all patches being masked. On one hand, this increases the difficulty of the pretext task to avoid the model learning trivial things such as recording environment for prediction; on the other hand, this also avoids building a memory bank of patches from different spectrograms and makes the algorithm less computationally intensive and less affected by the mini-batch size.

\section{Experiments}

\subsection{Pretraining Datasets}

In contrast to previous efforts that only use either speech dataset (e.g., in APC, wav2vec) or audio event dataset~\cite{saeed2021contrastive,niizumi2021byol}, in this work, we propose to use both speech and audio event datasets for pretraining to explore if the pretrained model can generalized to both speech and audio classification tasks. For both datasets, we only use the audio data and abandon the labels for self-supervised pretraining. 

\subsubsection{AudioSet-2M}
We use the AudioSet full training set (AudioSet-2M)~\cite{gemmeke2017audio} as our audio pretraining dataset. AudioSet is a multi-label audio event classification dataset that contains 2 million 10-second audio clips excised from YouTube videos with 527 sound classes including human sounds, animal sounds, sounds of things, music, natural sounds, environment sounds etc. It is worth mentioning that while about half of AudioSet-2M audio clips contain speech, speech might only appear in a small part of each clip as most AudioSet clips contain more than one sound. Therefore, AudioSet potentially does not have good coverage of speech and might not be sufficient to pretrain a good model for downstream speech tasks. 

\subsubsection{Librispeech}
In order to improve the coverage of speech data, we further use the Librispeech~\cite{panayotov2015librispeech} 960-hour training set as our speech pretraining dataset. Librispeech contains public domain audio books data in English, read by over 1,000 speakers, and is commonly used to train and evaluate speech recognition systems.

For both AudioSet and Librispeech data, we cut or pad each waveform to 10sec. We use 1,953k AudioSet samples and 281k Librispeech samples, and a total of 2,234k samples. We mix and shuffle the two datasets during pretraining.

\begin{figure}[!t]
    \centering
    \includegraphics[width=6.7cm]{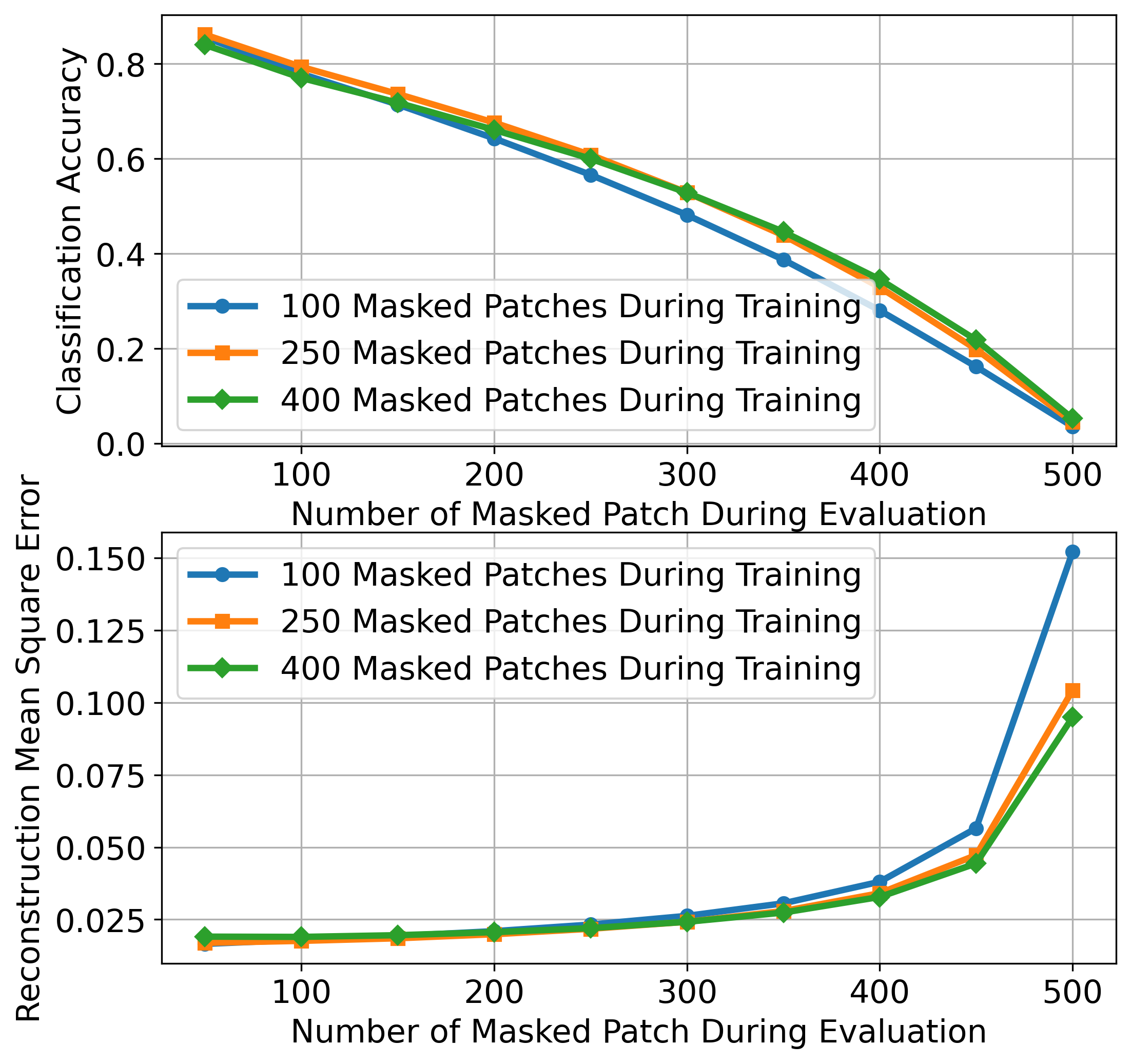}
    \caption{Prediction accuracy (upper) and reconstruction MSE (lower) of the masked patch modeling pretext tasks. We pretrain three AST models with a fixed number of 100, 250, and 400 masked patches, respectively, and evaluate their classification and reconstruction performance with various masked patch numbers from 50 to 500 on the validation set. While the AST model is pretrained with a fixed number of masked patches, we find it can perform well with a different number of masked patches in inference. As expected, the performance of the model drops with the increase of the number of masked patches, e.g., the AST models achieve over 80\% accuracy when the evaluation masked patch number is 50, but only around 35\% when the evaluation masked patch number is 400. This indicates the pretext tasks are neither trivial nor impossible.}
    \label{fig:upstream}
\end{figure}

\subsection{Performance of Pretext Tasks}
For pretraining the AST, we use a batch size of 24, an initial learning rate of 1e-4, and cut the learning rate into half if the pretext task performance on the validation set stops improving for 8k iterations. We optimize the network using the Adam optimizer~\cite{kingma2015adam}. We train the model for up to 800k iterations ($\sim$8.5 epochs). We tested different numbers of masked patches of 100, 250, and 400. We pretrain SSAST on 4$\times$ NVIDIA GTX Titan X or GTX Titan X Pascal GPUs, the pretraining takes about 10 days.

We show the masked spectrogram patch modeling performance in Figure~\ref{fig:upstream}. While the AST model is pretrained with a fixed number of masked patches, we find it can perform well with a different number of masked patches during inference. As expected, the performance of the model drops with the increase of the number of masked patches, e.g., the AST models achieve over 80\% accuracy when the evaluation masked patch number is 50, but only around 35\% when the evaluation masked patch number is 400, indicating the pretext tasks are neither trivial nor impossible. In general, the model pretrained with more masked patches performs better on the pretext tasks.

\subsection{Downstream Tasks and Datasets}
We evaluate the pretrained model on 6 commonly used audio and speech benchmarks. We use the same three benchmarks (AudioSet-20K, ESC-50, and Speech Commands V2) that the original AST has been tested on and use exactly the same setting intentionally to make a fair comparison. To further evaluate the model performance on downstream speech tasks and compare with previous self-supervised models that focus on speech, we test the pretrained AST on three additional benchmark Speech Commands V1, VoxCeleb 1, and IEMOCAP for keyword spotting, speaker identification, and emotion recognition, respectively. We report mean Average Precision (mAP) for the AudioSet-20K task and accuracy for all other tasks.

\subsubsection{AudioSet-20K (AS)} We use the AudioSet balanced training set and evaluation set for the multi-label audio event classification task. The AudioSet-20K training set is a class-balanced subset of AudioSet-2M that contains 20,785 audios. We test the model on the AudioSet evaluation set, which is disjoint with AudioSet-20K and AudioSet-2M. 

\subsubsection{ESC-50 (ESC)} We use the ESC-50 dataset~\cite{piczak2015esc} for the single-label audio event classification task. ESC-50 is an audio classification dataset consists of 2,000 5-second environmental audio recordings organized into 50 classes. 

\subsubsection{Speech Commands V2 (KS2)}
We use the Speech Commands V2~\cite{warden2018speech} for the keyword spotting task. The Speech Command V2 dataset consists of 105,829 1-second recordings of 35 common speech commands. 

\subsubsection{Speech Commands V1 (KS1)}
We also use the Speech Commands V1~\cite{warden2018speech} for the keyword spotting task, which is similar to Speech Commands V2, but only contains 10 classes of keywords, 1 class of silence, and an unknown class to include the false positive. 

\subsubsection{VoxCeleb 1 (SID)} 
We use the VoxCeleb 1 dataset~\cite{nagrani2020voxceleb} that contains speech from 1,251 speakers for the speaker identification task. The task goal is to classify each utterance by its speaker identity where speakers are in the same predefined set for both training and testing. 

\subsubsection{IEMOCAP (ER)} We use the IEMOCAP dataset~\cite{busso2008iemocap} that contains about 12 hours of emotional speech for the speech based emotion recognition task.

\subsection{Downstream Fine-tuning Details}
\label{sec:ftsetting}

To make a fair comparison with previous work, for the AudioSet-20K, ESC-50, and Speech Commands V2 experiments, we train and evaluate the model using the exact same training and evaluation settings with the original AST. Specifically, we use mixup training~\cite{tokozume2018learning}, SpecAugment~\cite{park2019specaugment}, an initial learning rate of 5e-5, 1e-4, and 2.5e-4 and train the model with 25, 50, and 30 epochs for AudioSet-20K, ESC-50, Speech Commands V2, respectively.

For the three benchmarks Speech Commands V1, VoxCeleb1, and IEMOCAP that the original AST has not been tested on, we use the standard SUPERB~\cite{yang2021superb} training and testing framework. Specifically, we search the learning rate from 1e-5 to 1e-3 for out SSAST model and all baseline models and train the model for up to 20k and 40k iterations for Speech Commands V1 and VocCeleb1, respectively. We use a fixed learning rate of 1e-4 and max iteration of 10k for IEMOCAP. Please refer to the AST and SUPERB papers for more details. For all downstream experiments, we use the \emph{end-to-end fine-tuning} setting, i.e., we do not freeze any layer of the pretrained AST.

For supervised pretrained models, we use the output of \texttt{[CLS]} token as the audio clip representation because supervision is given to the output of \texttt{[CLS]} in pretraining while we use mean pooling for self-supervised models as supervision is given to individual token in pretraining, keeping pretraining and fine-tuning consistent can slightly improve the performance and make the comparison fairer.

\subsection{Performance on Downstream Tasks}

We compare the following models in our experiments:

\begin{enumerate}
    \item \textbf{AST-Scratch}: AST model with appropriate initialization but without any pretraining.
    \item \textbf{AST-IM+KD}: AST model with supervised ImageNet pretraining, proposed in~\cite{gong21b_interspeech}. The model is pretrained with the ImageNet 2012 dataset in a supervised manner. In addition, during ImageNet pretraining, knowledge distillation from another convolution neural network is applied, which can noticeably improve the performance~\cite{touvron2020deit}. This is a strong baseline that achieves state-of-the-art results on AudioSet-20K, ESC-50, and Speech Commands V2.
    \item \textbf{AST-AudioSet}: AST model with supervised AudioSet-2M pretraining on the audio event classification task.
    \item \textbf{SSAST 250}: The proposed self-supervised AST model pretrained with 250 masked patches.
    \item \textbf{SSAST 400}: The proposed self-supervised AST model pretrained with 400 masked patches.
\end{enumerate}

\begin{table}[!t]
\caption{Comparison of self-supervised AST with baseline models on various benchmarks.}
\label{tab:mainres}
\centering
\begin{tabular}{@{}lcccccc@{}}
\toprule
\multicolumn{1}{c}{Model} & \multicolumn{6}{c}{Task}                                                                      \\ \midrule
\multicolumn{1}{c}{}      & AS            & ESC           & KS2           & KS1           & SID           & ER            \\ \midrule
AST-Scratch               & 14.8          & 41.9          & 92.6          & 87.2          & 30.1          & 51.9          \\ \midrule
\multicolumn{7}{l}{\textbf{Supervised Pretraining Baselines}}                                                             \\
AST-IM + KD               & \textbf{34.7} & 88.7          & 98.1          & 95.5          & 41.1          & 56.0          \\
AST-AudioSet              & 28.6          & 86.8          & 96.2          & 91.6          & 35.2          & 51.9          \\ \midrule
\multicolumn{7}{l}{\textbf{Proposed Self-Supervised AST}}  \\
SSAST 250               & 30.4          & 86.7          & \textbf{98.1} & \textbf{96.2} & \textbf{66.6} & 57.1          \\
SSAST 400               & 31.0          & \textbf{88.8} & 98.0          & 96.0          & 64.2          & \textbf{59.6}          \\ \bottomrule
\end{tabular}
\end{table}

\begin{figure}[!t]
    \centering
    \includegraphics[width=8.0cm]{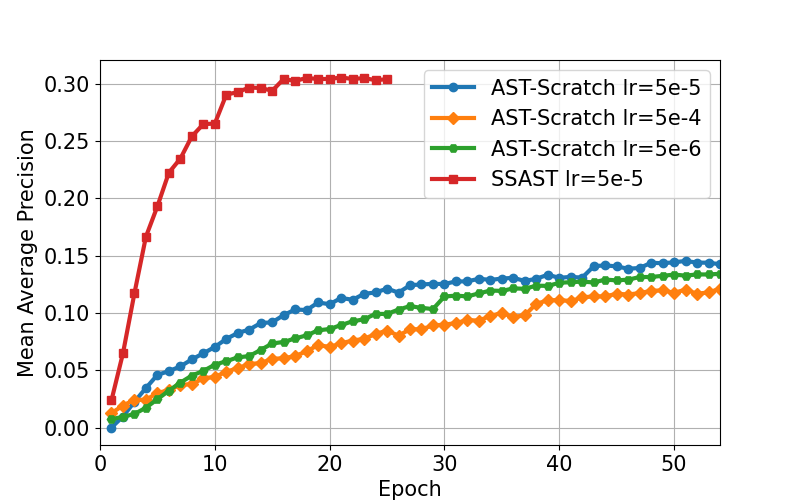}
    \caption{Comparing learning curves of AST trained from scratch and self-supervised AST on the AudioSet-20K task. The self-supervised framework helps AST train faster and better. Using a different learning rates, or  increasing training epochs does not improve the AST-scratch performance.}
    \label{fig:lr_curve}
\end{figure}

As shown in Table~\ref{tab:mainres}, we evaluate the above-mentioned 7 models on 6 benchmarks. Key findings include: First, the proposed self-supervised training framework can significantly boost the performance of AST with an average improvement of 60.9\%, e.g., SSAST achieves 0.310 mAP on the AudioSet-20K while AST-Scratch only achieves 0.148 mAP. As shown in Figure~\ref{fig:lr_curve}, the proposed self-supervised framework helps AST train faster and better. Further, the improvement is consistent over all audio and speech benchmarks, demonstrating the proposed self-supervised training framework is effective and generalizable. Second, AudioSet-2M supervised pretraining is quite strong for audio event classification tasks (AS and ESC) that are in the same domain with AudioSet, but performs poorly on speech tasks, showing the limitation of supervised pretraining. Surprisingly, cross-domain supervised ImageNet pretraining with knowledge distillation performs quite well on all tasks, and still achieves the best performance on the AudioSet-20K task. Third, even when compared with strong supervised baselines, the proposed SSAST models still get the best results on all benchmarks except AS, showing the proposed self-supervised model potentially can be used as a powerful generic audio classifier. 

\begin{table}[!t]
\caption{Ablation study on the impact of number of masked patches, pretext task, and pretraining data.}
\label{tab:ablation}
\begin{tabular}{@{}lcccccc@{}}
\toprule
\multicolumn{1}{c}{Setting}                                                             & \multicolumn{6}{c}{Task}                                                                                                                                                          \\ \midrule
\multicolumn{1}{c}{}                                                                    & AS                          & ESC                         & KS2                         & KS1                         & SID                         & ER                          \\ \midrule
From Scratch                                                                            & 14.8                        & 41.9                        & 92.6                        & 87.2                        & 30.1                        & 51.9                        \\ \midrule
\multicolumn{7}{l}{\textbf{\# Masked Patches}}                                                                                                                                                                                                                              \\
100                                                                                     & 28.7                        & 85.3                        & 98.0                        & 94.9                        & 62.1                        & 57.3                        \\
250                                                                                     & 30.4                        & 86.7                        & \textbf{98.1}               & \textbf{96.2}               & \textbf{66.6}               & 57.1                        \\
400 (Default)                                                                              & \textbf{31.0}               & \textbf{88.8}               & 98.0                        & 96.0                        & 64.3                        & \textbf{59.6}               \\ \midrule
\multicolumn{7}{l}{\textbf{Pretext Task}}                                                                                                                                                                                                                                   \\
Discriminative                                                                          & 30.6                        & 85.6                        & 98.0                        & 94.2                        & 61.4                        & 57.5                        \\
Generative                                                                              & 16.1                        & 74.2                        & 96.6                        & 93.3                        & 40.1                        & 54.3                        \\
Joint (Default)                                                                            & \textbf{31.0}               & \textbf{88.8}               & \textbf{98.0}               & \textbf{96.0}               & \textbf{64.3}               & \textbf{59.6}               \\ \midrule
\multicolumn{7}{l}{\textbf{Pretraining Data}}                                                                                                                                                                                                                                  \\
AudioSet-20K                                                                            & 25.7                        & 82.2                        & 97.6                        & 93.8                        & 43.8                        & 55.4                        \\
AudioSet 2M                                                                             & 29.0                        & 84.7                        & 97.8                        & 94.8                        & 57.1                        & 56.8                        \\
{\color[HTML]{656565} \begin{tabular}[c]{@{}l@{}}AudioSet 2M\\ Supervised\end{tabular}} & {\color[HTML]{656565} 28.6} & {\color[HTML]{656565} 86.8} & {\color[HTML]{656565} 96.2} & {\color[HTML]{656565} 91.6} & {\color[HTML]{656565} 35.2} & {\color[HTML]{656565} 51.9} \\
Librispeech                                                                             & 22.9                        & 80.0                        & 97.8                        & 95.6                        & 60.8                        & 58.3                        \\
Joint (Default)                                                                            & \textbf{31.0}               & \textbf{88.8}               & \textbf{98.0}               & \textbf{96.0}               & \textbf{64.3}               & \textbf{59.6}               \\ \bottomrule
\end{tabular}
\end{table}

\subsection{Performance Impact of Pretraining Settings}

We set the AST pretrained with 400 masked patches, joint discriminative and generative objectives, on both AudioSet-2M and Librispeech as the base model. We then change one factor at a time to observe the performance impact. 

\subsubsection{Impact of the Number of Masked Patches}
As shown in Table~\ref{tab:ablation}, upper section, we find masking 100 patches is too simple a task, and leads to the worst performance for all downstream tasks. Masking 400 patches leads to better performance on audio event classification tasks, while masking 250 patches leads to better performance on speech tasks, but the overall performance is similar. 

\subsubsection{Impact of Pretext Tasks}
As shown in Table~\ref{tab:ablation}, middle section, we find that a discriminative objective leads to better performance than the generative objective for all tasks, but joint discriminative and generative objective always achieves the best performance, indicating that the discriminative and generative objectives are complementary.

\subsubsection{Impact of Pretraining Data}

We pretrain the AST model using 1) AudioSet-20K, 2) AudioSet-2M only, 3) Librispeech only, and 4) both AudioSet-2M and Librispeech, and compare the performance of the pretrained models on the downstream tasks. As shown in Table~\ref{tab:ablation}, bottom section, we have the following key findings: First, increasing the pretraining data volume improves the performance of downstream tasks, e.g., AudioSet-2M pretrained model always outperforms AudioSet-20K pretrained model, but the proposed self-supervised framework can still noticeably improve the AST model with limited pretraining data, e.g., when pretrained and fine-tuned on the same AudioSet-20K data, the proposed SSAST model achieves 0.257 mAP, and significantly outperforms the AST-Scratch model. Second, with the same AudioSet-2M pretraining data, the proposed self-supervised framework leads to similar or even better results compared with the supervised pretraining method, particularly for the speech tasks, showing that the proposed self-supervised framework is more generalizable. Third, as expected, a model pretrained with AudioSet-2M is better for audio classification and a model pretrained with Librispeech is better for speech tasks, but training with both sets always leads to the best results, showing that it is beneficial to combine pretraining datasets in audio and speech domains.

\begin{figure}[!t]
    \centering
    \includegraphics[width=7.9cm]{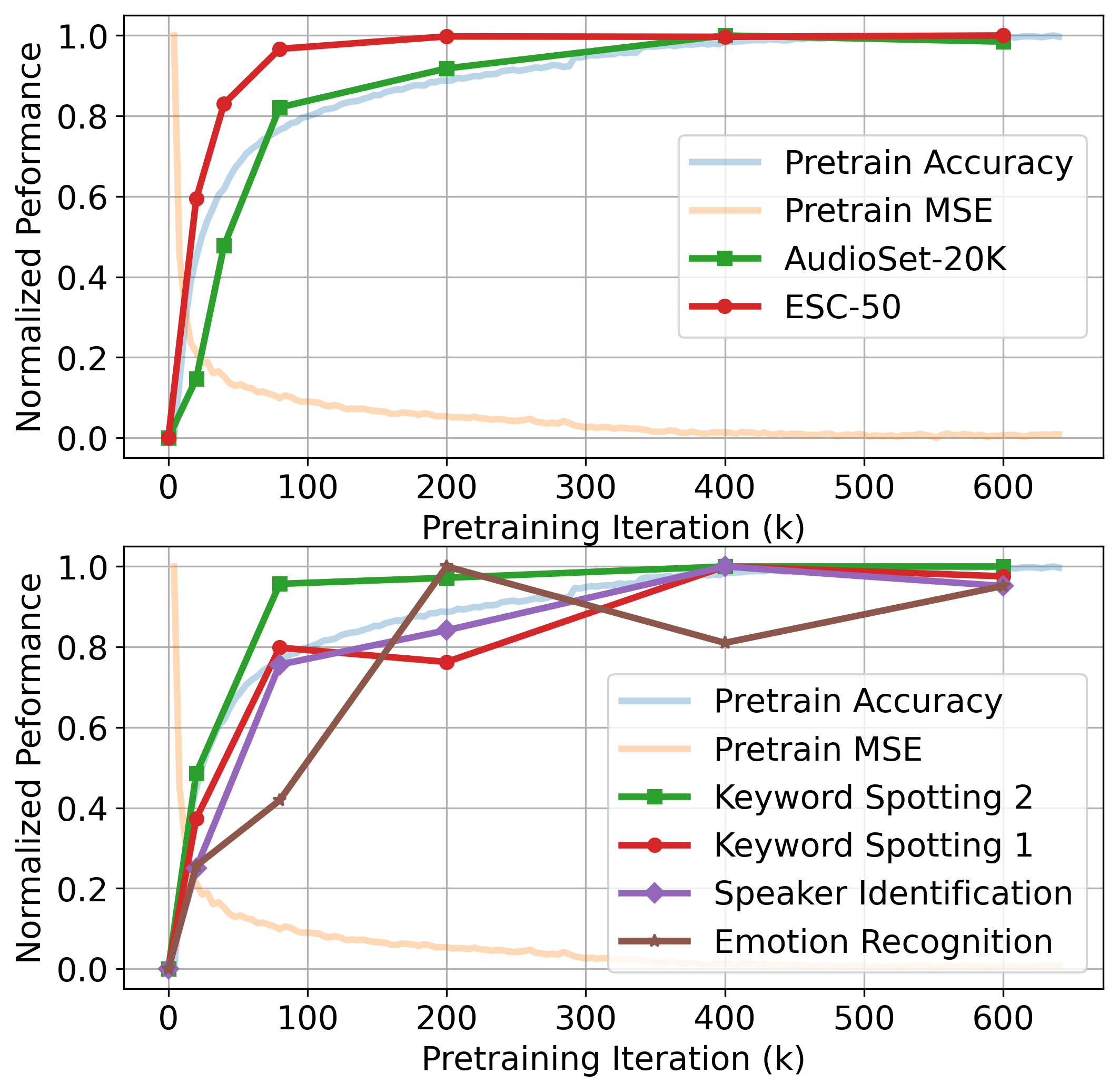}
    \caption{Performance correlation between pretraining tasks and downstream tasks (upper: audio classification tasks, lower: speech tasks). We save the checkpoint models at iteration 20, 40, 80, 200, 400, and 600 during pretraining, then fine-tune and evaluate these checkpoint models on the downstream tasks. For better visualization, we normalize the performance of each task in the range $[0,1]$. We observe that the model pretrained with more iterations generally performs better on downstream tasks, which further confirms that the pretraining pretext tasks can benefit all downstream tasks.}
    \label{fig:corrupdown}
\end{figure}

\subsubsection{Performance Correlation between Pretraining and Downstream Tasks}
We save the checkpoint models at iteration 20, 40, 80, 200, 400, and 600 during pretraining, then fine-tune and evaluate these checkpoint models on the downstream tasks. We observe the performance of pretraining tasks and downstream tasks are highly correlated, i.e., the model pretrained with more iterations generally performs better on downstream tasks, which further confirms that the pretraining pretext tasks benefit all downstream tasks.

\subsection{Performance Impact of AST Model Size}

\begin{table}[]
\caption{Comparison of AST model of different sizes ($^*$ use larger learning rate for the last linear classification layer).}
\setlength\tabcolsep{5pt}
\centering
\label{tab:mdlsize}
\begin{tabular}{@{}lcccccc@{}}
\toprule
\multicolumn{1}{c}{Model} & \multicolumn{6}{c}{Task}                \\ \midrule
                          & AS   & ESC  & KS2  & KS1  & SID  & ER   \\ \midrule
Tiny-Scratch              & 15.1 & 34.8 & 92.4 & 87.7 & 24.2 & 50.8 \\
Tiny-SSAST                & 27.1$^*$ & 79.5 & 97.2 & 94.8 & 55.1 & 55.7 \\ \midrule
Small-Scratch             & 16.5 & 37.8 & 93.3 & 87.4 & 23.8 & 51.2 \\
Small-SSAST               & 30.8$^*$ & 85.4 & 97.7 & 95.4 & 60.9 & 58.7 \\ \midrule
Base-Scratch              & 14.8 & 41.9 & 92.6 & 87.2 & 30.1 & 51.9 \\
Base-SSAST                & 31.0 & 88.8 & 98.0 & 96.0 & 64.2 & 59.6 \\ \bottomrule
\end{tabular}
\end{table}

In all previous experiments, we use the original AST~\cite{gong21b_interspeech} architecture to make a direct performance comparison. We refer to this model as the \emph{base} AST model. In this section, we further test the following AST architectures to study the impact of model size.

\begin{enumerate}
    \item \textbf{Tiny Model}: The Transformer encoder has 12 layers with 3 attention heads and an embedding dimension of 192. The tiny model has 6M parameters.
    \item \textbf{Small Model}: The Transformer encoder has 12 layers with 6 attention heads and an embedding dimension of 384. The small model has 23M parameters. 
    \item \textbf{Base Model}: The model described in Section~\ref{sec:arc} that is used as the default model throughout the paper. The Transformer encoder has 12 layers with 12 attention heads and an embedding dimension of 768. The base model has 89M parameters.
\end{enumerate}

For each model architecture, we compare the performance of the from-scratch model and the self-supervised pretrained SSAST model (pretrained with 400 masked patches) and show the results in Table~\ref{tab:mdlsize}. Key findings are as follows:

First, the MSPM self-supervised pretraining consistently enhances the performance of all three model architectures, showing that MSPM is model size agnostic. Small models that are unlikely to be over-parameterized also get performance improvement with MSPM pretraining. 

Second, when trained from scratch, the larger AST model does not always get the best performance, e.g., the small AST model outperforms the base AST model on AS, KS1, and KS2 tasks. This is as expected since larger models are harder to train with limited data. However, we find that with MSPM self-supervised pretraining, larger AST models always perform better, demonstrating that MSPM can unlock the potential of models with higher capacity. This also suggests that further scaling up the base AST model can potentially achieve even better performance.

We also observe that using a larger learning rate for the last linear layer during fine-tuning improves the performance for tiny and small SSAST models on the AS task, e.g., for small SSAST model, using a learning rate of 5e-3 for the last linear layer and 5e-5 for all other layers leads to an mAP of 0.308 while using a learning rate of 5e-5 for the entire model leads to an mAP of only 0.272. Nevertheless, we find this trick is only useful for tiny and small self-supervised pretrained models for some downstream tasks, it does not improve the performance of from-scratch models.

\subsection{Comparing Patch-based and Frame-based AST}
\label{sec:patchvsframe}

In all previous experiments, we follow the original AST~\cite{gong21b_interspeech} to split the audio spectrogram into $16\times16$ square patches. In~\cite{gong21b_interspeech}, it was found that splitting the spectrogram into frame-like rectangle patches in the temporal order leads to better performance when the model is trained from scratch. However, ImageNet supervised pretrained model performs significantly better than the from-scratch model, which also constrains the original AST to use square patches. In contrast, our proposed MSPM self-supervised pretraining supports any patch size and shape including a conventional frame. As discussed in Section~\ref{sec:method}, heuristically, square patch based pretraining could capture correlation in frequency bands in addition to time frames, which is potentially useful when the input has a complex frequency structure (e.g., natural sounds). For clarity, we refer to the AST model that uses square patches and frame-like rectangle patches as patch-based AST model and frame-based AST model, respectively. In this section, we compare patch-based and frame-based AST models in both from-scratch setting and self-supervised pretraining setting. Specifically, the two models have exactly the same architecture except the patch splitting layer, for the patch-based AST model, we use $16\times16$ patches as described in Section~\ref{sec:method}; for the frame-based AST model, instead of splitting the spectrogram into $16\times16$ patches, we split the spectrogram into $128\times2$ patches in the temporal order (128 is the number of frequency bins of the spectrogram). Patches are split without overlap during pretraining and are split with an overlap of 1 on the time dimension during fine-tuning. This makes a fair comparison as the area of the patch is the same and the number of patches after splitting is similar. In the pretraining setting, both models are pretrained using the method described in Section~\ref{sec:method}. The only pretraining setting difference is that we do not cluster the masked frames for frame-based AST because this would lower the pretext and downstream task performance, instead, we just random sample the masked frame for frame-based AST pretraining. We test models pretrained with 250 and 400 masked patches (frames) and show the results in Table~\ref{tab:patchvsframe}. Key findings are as follows:

First, when trained from scratch, frame-based AST always performs better than patch-based AST (except ER), which is consistent with the finding in~\cite{gong21b_interspeech} and as expected because 1-D temporal structure is easier to learn than 2-D temporal-frequency structure. Second, after MSPM self-supervised pretraining, frame-based AST still outperforms patch-based AST on speech tasks (KS1, KS2, SID, and ER) but the advantage becomes much smaller. Patch-based AST performs better on audio tasks (AS and ESC). MSPM significantly improves the performance of both patch-based and frame-based AST, but the improvement is noticeably larger for patch-based AST (except ER), which verifies our hypothesis that square patch based pretraining can be more effective, particularly for data that has a complex frequency structure such as natural sounds. Our experiment also demonstrates that MPSM is patch shape agnostic, it also works well with frame-based AST and makes frame-based SSAST a strong model for speech tasks. In contrast, previous ImageNet pretraining only supports square patches.


\begin{table}[t]
\caption{Comparison of frame and patch based AST models.}
\label{tab:patchvsframe}
\setlength\tabcolsep{3.2pt}
\centering
\begin{tabular}{llccccc}
\toprule
\multicolumn{1}{c}{Model} & \multicolumn{6}{c}{Task}                                                                      \\ \midrule
                          & AS            & ESC           & KS2           & KS1           & SID           & ER            \\ \midrule
Frame-Scratch             & \textbf{16.6} & \textbf{53.7} & \textbf{96.0} & \textbf{91.7} & \textbf{54.9} & 51.2          \\
Patch-Scratch             & 14.8          & 41.9          & 92.6          & 87.2          & 30.1          & \textbf{51.9} \\ \midrule
SSAST-Frame-250           & 27.1          & 84.0          & 98.0          & \textbf{96.6} & \textbf{73.6} & \textbf{58.3} \\
SSAST-Patch-250           & \textbf{30.4} & \textbf{86.7} & \textbf{98.1} & 96.2          & 66.6          & 57.1          \\ \midrule
SSAST-Frame-400           & 29.2          & 85.9          & \textbf{98.1} & \textbf{96.7} & \textbf{80.8} & \textbf{60.5} \\
SSAST-Patch-400           & \textbf{31.0} & \textbf{88.8} & 98.0          & 96.0          & 64.2          & 59.6          \\ \midrule \midrule
Frame-Improvement         & 12.6          & 32.2          & 2.1           & 5.0           & 25.9          & \textbf{9.3}  \\
Patch-Improvement         & \textbf{16.2} & \textbf{46.9} & \textbf{5.4}  & \textbf{8.8}  & \textbf{34.1} & 7.7           \\ \bottomrule
\end{tabular}
\end{table}

\subsection{Comparing with Existing Speech Self-supervised Pretraining Frameworks}

Finally, we compare the performance of SSAST with existing speech self-supervised pretraining frameworks. Since these frameworks are designed for speech tasks and are only pretrained on speech datasets, we only compare with them on the speech benchmarks. Specifically, we compare three SSAST models with previous models: 1) \textbf{SSAST-Patch (Librispeech)}: Patch-based SSAST model pretrained on only Librispeech (same pretraining data with previous speech self-supervised models); 2) \textbf{SSAST-Patch}: Patch-based SSAST model pretrained on both AudioSet and Librispeech; and 3) \textbf{SSAST-Frame} SSAST model described in Section~\ref{sec:patchvsframe} that uses frame-like patches and is  pretrained on both AudioSet and Librispeech.

\begin{table}[t]
\caption{Comparison of SSAST and existing speech self-supervised pretraining frameworks ($^*$frozen setting results).}
\label{tab:comparesota}
\centering
\begin{tabular}{@{}lccc@{}}
\toprule
\multicolumn{1}{c}{Model}          & \multicolumn{3}{c}{Task}                                                                \\ \midrule
                                   & KS1                         & SID                         & ER                          \\ \midrule
APC~\cite{chung2019unsupervised}                            & 94.0                        & 60.4                        & 59.3                        \\
Wav2vec~\cite{schneider2019wav2vec}                            & 96.2                        & 56.6                        & 59.8                        \\
{\color[HTML]{656565} Wav2vec 2.0~\cite{baevski2020wav2vec}$^*$} & {\color[HTML]{656565} 96.2} & {\color[HTML]{656565} 75.2} & {\color[HTML]{656565} 63.4} \\
{\color[HTML]{656565} HuBERT~\cite{hsu2021hubert}$^*$}     & {\color[HTML]{656565} 96.3} & {\color[HTML]{656565} 81.4} & {\color[HTML]{656565} 64.9} \\ \midrule
SSAST-Patch (Librispeech Only)     & 95.6                        & 60.8                        & 58.3                        \\
SSAST-Patch                        & 96.0                        & 64.3                        & 59.6                        \\
SSAST-Frame                        & 96.7                        & 80.8                        & 60.5                        \\ \bottomrule
\end{tabular}
\end{table}

\subsubsection{Comparing with APC and wav2vec 1.0}

We first compare SSAST models with autoregressive predictive coding (APC)~\cite{chung2019unsupervised}, a generative pretraining framework, and wav2vec 1.0~\cite{schneider2019wav2vec}, a discriminative pretraining framework. We evaluate APC and wav2vec 1.0 in both fine-tuned and frozen settings and report the best result. As shown in Table~\ref{tab:comparesota}, SSAST models match or outperform APC and wav2vec 1.0 on all three benchmarks.

\subsubsection{Comparing with wav2vec 2.0 and HuBERT}

We then compare SSAST models with the state-of-the-art wav2vec 2.0~\cite{baevski2020wav2vec} and HuBERT~\cite{hsu2021hubert} models. Specifically, we compare the base model that is pretrained on Librispeech 960 dataset. Due to the complexity of finding optimal hyperparameters and the large computation cost for fine-tuning these two models, we only report the results in the frozen setting. As shown in Table~\ref{tab:comparesota}, frozen wav2vec and HuBERT can already match or outperform fine-tuned SSAST for speech tasks. Nevertheless, it is worth noting that although wav2vec 2.0 and HuBERT perform better, they are pre-trained with 64/32 GPUs and hence have larger batch sizes than our SSAST that is trained with 4 GPUs. The computational resource difference could greatly impact the performance, e.g., for HuBERT, using 8 GPUs leads to 40\% WER while 32 GPUs leads to below 20\% WER. With more computational resources and larger batch size, SSAST potentially can achieve better results.

\section{Related Work}

\subsubsection{Pure Transformer Based Models} Self-attention models, especially the Transformer~\cite{vaswani2017attention}, have been widely used in natural language processing. Recently, pure Transformer models, e.g., Vision Transformer~\cite{dosovitskiy2021image,touvron2020deit,yuan2021tokens} and Audio Spectrogram Transformer~\cite{gong21b_interspeech}, are found to outperform CNN based models for vision tasks and audio classification. Such models differ from CNN models or CNN-Attention hybrid models in that they do not contain non-degenerated convolutions~\cite{chen2021empirical} and have less inductive bias such as spatial locality and translation equivariance. However, it is found that such pure Transformer models require a lot of training data to perform well~\cite{dosovitskiy2021image}.

\subsubsection{Self-Supervised Learning}

In the vision domain, self-supervised Vision Transformer has been studied in~\cite{caron2021emerging,chen2021empirical,atito2021sit}. In addition, patch based self-supervised framework has been extensively studied in the vision domain, e.g., in~\cite{noroozi2016unsupervised,trinh2019selfie,bao2021beit}. However, to the best of our knowledge, the self-supervised Audio Spectrogram Transformer and patch based self-supervised learning framework has not been studied in the audio and speech domain. Previous self-supervised learning frameworks in the speech domain are mainly based on CNN, RNN, or CNN-Transformer hybrid models with the pretext task of predicting past, current, or future frames~\cite{chung2019unsupervised,oord2018representation,liu2020mockingjay,schneider2019wav2vec}. In contrast, the proposed MSPM framework allows the model to learn both the temporal and frequency structure of the spectrogram. Further, most previous research only focuses on learning either a speech or audio representation, only a few efforts~\cite{saeed2021contrastive,niizumi2021byol} studied learning a general audio and speech representation. However, both efforts pretrain the model with only AudioSet. In contrast, we explore pretraining the AST model with both AudioSet and Librispeech. Finally, we pretrain the model with joint discriminative and generative objectives, which is also novel in the audio and speech domain and only has been explored in~\cite{pascual2019learning,jiang2020speech,ravanelli2020multi}.

\section{Conclusion}
This paper aims to reduce the need for large amounts of labeled data for the AST self-attention based audio and speech classification model by leveraging self-supervised learning. We propose MSPM, a novel patch-based joint discriminative and generative pretraining framework. In order to make the pretrained model generalize to both audio and speech tasks, we pretrain AST using both AudioSet and Librispeech, and evaluate on six downstream benchmarks including audio event classification, keyword spotting, speaker identification, and emotion recognition. 

With extensive experiments, we observe the following key findings. First, the proposed MSPM self-supervised pretraining framework significantly improves the performance of AST for all downstream tasks with an average improvement of 60.9\%. Our SSAST model can match or even outperform previous supervised pretrained models and shows better generalization capability, indicating that the proposed MSPM can replace supervised pretraining that requires a large amount of labeled data. Second, we find that pretraining the model with both generative and discriminative objectives leads to a better performance than using a single objective, similarly, pretraining the model on both speech and audio datasets leads to better performance than using data from a single domain. Third, the flexibility of MSPM on patch shape allows us to explore frame-based AST. We find that frame-based AST always outperforms patch-based AST in the from-scratch setting, but patch-based pretraining leads to a larger improvement from the random-initialized models. After MSPM pretraining, the patch-based AST wins on the audio tasks while the frame-based AST wins on the speech tasks. We plan to investigate the reason for this difference in our future work. Finally, we find MSPM allows us to scale up the AST model, with MSPM pretraining, larger AST always performs better. In contrast, in the from-scratch setting, scaling up the model may cause a performance drop. Nevertheless, the current version of SSAST is pretrained with a small batch size due to computational resource limitations. In the future, we plan to further investigate the scaling law of AST.

\section*{Acknowledgments}
We thank the anonymous reviewers for their insightful comments and suggestions. This work is partly supported by Signify.

\bibliography{aaai22}

\section*{Appendix}

\subsection*{Downstream Task Dataset and Evaluation Protocol Details}

We show more details about the downstream datasets and evaluation protocols.

\subsubsection{AudioSet-20K} We use the AudioSet balanced training set and evaluation set for the multi-label audio event classification task. The AudioSet-20K training set is a class-balanced subset of AudioSet-2M that contains 20,785 audio clips. For this task, we use the \emph{mean averaged precision (mAP)} as the matrix. We test the model on the AudioSet evaluation set, which is disjoint with both AudioSet-20K and AudioSet-2M. 

\subsubsection{ESC-50} We use the ESC-50 dataset~\cite{piczak2015esc} for single audio event classification task. ESC-50 is an audio classification dataset consists of 2,000 5-second environmental audio recordings organized into 50 classes. For this task, we follow the standard 5-fold cross-validation to evaluate our model and report the \emph{accuracy}. The difference between ESC-50 and AudioSet-20K is that each ESC-50 audio clip only contains a single event and the total data volume size is 10 times smaller than AudioSet-20K.

\subsubsection{Speech Commands V2-35}
We use the Speech Commands V2-35~\cite{warden2018speech} for the keyword spotting task. The Speech Command V2-35 dataset consists of 105,829 1-second recordings of 35 common speech commands. The training, validation, and test set contains 84,843, 9,981, and 11,005 samples, respectively. We fine-tune the pretrained model on the training set, select the model using the validation set, and report the \emph{accuracy} on the test set. 

\subsubsection{Speech Commands V1}
We also use the Speech Commands V1~\cite{warden2018speech} for the keyword spotting KS task, which is similar to Speech Commands V2, but only contains 10 classes of keywords, 1 class of silence, and an unknown class to include the false positive. For Speech Commands V1, we use the SUPERB evaluation framework~\cite{yang2021superb} and report the \emph{accuracy} on the test set.

\subsubsection{VoxCeleb 1} 
We use VoxCeleb 1 dataset~\cite{nagrani2020voxceleb} for the speaker identification task. The VoxCeleb 1 dataset contains 352 hours of speech from 1,251 speakers. The goal of this task is to classify each utterance for its speaker identity where speakers are in the same predefined set for both training and testing. For VoxCeleb 1, we use the SUPERB evaluation framework~\cite{yang2021superb} and report the \emph{accuracy} on the test set.

\subsubsection{IEMOCAP} We use the IEMOCAP dataset~\cite{busso2008iemocap} for the speech based emotion recognition (ER) task that contains about 12 hours of emotional speech from 10 speakers. For IEMOCAP, we use the SUPERB evaluation framework~\cite{yang2021superb}, which follows the conventional IEMOCAP evaluation protocol that drops the unbalance emotion classes to leave the four roughly balanced classes (neutral, happy, sad, angry), conduct 5-fold cross-validation, and report the mean \emph{accuracy}.

\subsection*{Experiment Details}

\subsubsection{Utterance Representation} For the reason mentioned in Section 2.1 of the main manuscript, for supervised pretrained baseline (\textbf{AST-IM+KD} and \textbf{AST-AudioSet}), we use the output representation of the \texttt{[CLS]} token as the utterance representation because supervision is given to the output of \texttt{[CLS]} in pretraining while we use the average patch representation of all patches as the utterance representation (i.e., applying a mean pooling over $O$) for \textbf{AST-Scratch} and \textbf{SSAST} because supervision is given to individual token in pretraining, keeping pretraining and fine-tuning settings consistent can slightly improve the performance and make the comparison fairer. Model without pretraining might need more epochs to train, therefore, we train AST-Scratch models with more epochs until convergence for a fair comparison.

\subsubsection{Model Initialization}
We use default \texttt{PyTorch} initialization for weights for all layers for the AST models, which is the same as the original AST paper~\cite{gong21b_interspeech}.

\subsection*{Computing Software and Infrastructure}
We run our experiments with \texttt{Python 3.7.4}, \texttt{PyTorch 1.9.0}, and \texttt{CUDA 10.2}. We run our experiments on 4$\times$ NVIDIA GTX Titan X or GTX Titan X Pascal GPUs; Intel Xeon CPU E5-2623 v3 or E5-2620 v4 CPUs, and 128GB memory.

\end{document}